
\input amstex
\documentstyle{amsppt}
\magnification=1200
\NoBlackBoxes
\define\proj{\operatorname{proj}}
\define\sLtwo{\operatorname{sl}(2,\Bbb C)}
\define\Proj{\operatorname{Proj}}
\document
\subhead ALGEBRAIC STRUCTURES OF QUANTUM PROJECTIVE FIELD THEORY RELATED TO
FUSION
AND BRAIDING.\newline HIDDEN ADDITIVE WEIGHT\endsubhead

\

\

\ \newline\rm Denis Juriev\footnote{On leave from Mathematical Division,
Research
Institute for System Studies, Russian Academy of Sciences, Moscow, Russia (e
mail: juriev\@systud.msk.su).}\newline

\

\ \newline Laboratoire de Physique Th\'eorique de l'\'Ecole Normale
Sup\'erieure, 24 rue
Lhomond, 75231 Paris Cedex 05, France\footnote{Unit\'e Propre du Centre
National de la Recherche Scientifique associ\'ee \`a l'\'Ecole Normale
Sup\'erieure et \`a l'Universit\'e de Paris--Sud.}\newline
E mail: juriev\@physique.ens.fr

\

\

\

\

\

The interaction of various algebraic structures, describing fusion, braiding
and group symmetries in quantum projective field theory, is an object of an
investigation in the paper. Structures of projective Zamolodchikov algebras,
their representations, spherical correlation functions, correlation characters
and envelopping QPFT--operator algebras, projective \"W--algebras, shift
algebras, infinite dimensional R--matrices $R_{\proj}(u)$ and $R^*_{\proj}(u)$
of the QPFT, braiding admissible QPFT--operator algebras and projective
G--hypermultiplets are explored.

It is proved (in the formalism of shift algebras) that $\sLtwo$--primary
fields are characterized by their projective weights and by the hidden
additive weight, a hidden quantum number discovered in the paper.

The special attention is paid to various constructions of projective
$G$--hypermul\-ti\-plets (QPFT--operator algebras with $G$--symmetries).

\

\

PASC Numbers:

\pagebreak

\ \newline

\subhead I. INTRODUCTION \endsubhead

\ \newline

In view of the fact that an appearance of the representation theory depends on
its applications, one might hope that a psychological change, which was
provoked by a development of modern quantum field theory, would open new
perspectives in the theory. Such hopes were confirmed by a lot of recent
papers$^{1-6}$. For example, accordingly to a program paper by G.Moore and
N.Seiberg$^6$ it is rather interesting to consider the concept of group being
supplemented by various new concepts related to fusion and braiding. They
considered some of those structures: vertex operator algebras, representations
of bordism categories and quantum groups as well as relations between them.

An appearing of this paper is motivated by the following arguments. As one
knows from the history of representation theory its primitive is a
finite--dimensional Lie algebra $\sLtwo$ more than infinite--dimensional
Kac--Moody and Virasoro algebras. So it is quite natural that the
braiding--fusion structures connected with $\sLtwo$ should attract a special
interest. In other words one should examine various algebraic structures of
the quantum projective ($\sLtwo$--invariant) field theory but not only ones of
the conformal or Kac--Moody--invariant field theories. Of course, such
examinations should be based on the general theoretical results, which were
obtained in papers on these theories.

These considerations explain the fact that structures of the quantum
projective field theory (QPFT), connected with fusion and braiding, are
investigated in this paper.

It is necassary to say some words about the tasks of the paper. Our task is
not to introduce principally new structures, which have not been known.
Moreover, the basic concepts of the paper are related to an operator product
expansion (OPE), so the alternative but deeply connected approaches, related
to quantum groups$^{2-4}$ and bordism categories$^5$, are not considered in
the paper. Our task is to examine the OPE--structures of the QPFT in
details,to describe some of them, which have no analogs in the conformal case,
to write some explicit formulas, whose analogs in the conformal case have not
be written in view of serious difficulties connected with the
infinite--dimensionality of the Virasoro and Kac--Moody algebras.

The structures of projective Zamolodchikov algebras, their representations,
spherical correlation functions and envelopping QPFT--operator algebras,
projective \"W--algebras, infinite dimensional R--matrices of the QPFT,
braiding admissible QPFT--operator algebras and projective
$G$--hypermultiplets are explored.

It is proved (in the formalism of shift algebras) that $\sLtwo$--primary
fields are characterized by their projective weights and by the hidden
additive weight, a hidden quantum number discovered in the paper.

The paper contains 11 theorems, which have an original character.
\newpage

\subhead II. FUSION IN THE QUANTUM PROJECTIVE FIELD THEORY: QPFT--OPERATOR
ALGEBRAS \endsubhead

\ \newline

\bf Definition 1\rm:$^{6-8}$ The QFT--operator algebra (operator algebra of
quantum field theory) is the pair $(H,t^k_{ij}(\vec x))$, where $H$ is a
linear space, $t^k_{ij}(\vec x)$ is the $H$--valued tensor field on $\Bbb C^n$
or $\Bbb R^n$ of the fixed analytic class such that

$$t^m_{ij}(\vec x-\vec y)t^l_{mk}(\vec y)=t^l_{im}(\vec x)t^m_{jk}(\vec y)\tag
1 $$

\bf Remark\rm : Let's introduce the operators $l_{\vec x}(e_i)$ such that

$$l_{\vec x}(e_i)e_j=t^k_{ij}(\vec x)e_k\tag 2 $$

then the following relations would hold

$$ l_{\vec x}(e_i)l_{\vec y}(e_j)=t^k_{ij}(\vec x-\vec y)l_{\vec
y}(e_k)\quad\text{(operator product expansion (OPE))}\tag 3$$

and

$$ l_{\vec x}(e_i)l_{\vec y}(e_j)=l_{\vec y}(l_{\vec x-\vec
y}(e_i)e_j)\quad\text{(duality).}\tag 4$$

The QFT--operator algebra turns into the ordinary associative algebra iff
$dt^k_{ij}(\vec x)=0$.

\bf Definition 2\rm:$^8$ The QFT--operator algebra $(H,t^k_{ij}(u),u\in\Bbb
C)$ is called the QPFT--operator algebra iff
\roster
\item"a)" $H$ is the direct sum of the direct integral of The Verma modules
$V_a$ over the Lie algebra $\sLtwo$ with the highest vectors $v_a$ and the
highest weights $h_a$;
\item"b)" the operator field $l_u(v_a)$ is primary, i.e.
$$ [L_k,l_u(v_a)]=(-u)^k(u\frac{\partial}{\partial u}+(k+1)h_a)l_u(v_a),$$
where $L_k$ ($k=-1,0,1$) are the generators of the Lie algebra $\sLtwo$:
$[L_k,L_j]=(k-j)L_{k+j}$;
\item"c)" the rule of secondaries' generation holds
$$ L_{-1}l_u(F)=l_u(L_{-1}F).$$
\endroster

\define\Mat{\operatorname{Mat}}
\define\Vrt{\operatorname{Vert}}
It maybe shown that QPFT--operator algebras are just the same as the
QPFT--subalgebras in $\Mat_N(\Vrt(\sLtwo))$, where $\Vrt(\sLtwo)$ is the
standard QPFT--operator algebra --- the algebra of the vertex operators for
$\sLtwo$. It seems to be convenient as well as necessary to present the
construction of the QPFT--operator algebra $\Vrt(\sLtwo))$ below. To do it we
need in a "model formalism".

\bf Definition 3\rm:$^8$ The model of the Verma modules over the Lie algebra
$\sLtwo$ is the representation of this algebra in the direct integral of the
Verma modules $V_h$ over $\sLtwo$.

The model admits several realizations. One of them is the
BGG--realization$^8$. The model space is the space of all holomorphic
functions of two complex variables $t$ and $z$, where $z$ belongs to the
complex plane $\Bbb C$ and $t$ belongs to the universal covering $\tilde{\Bbb
C}^*$ of the complex plane without zero $\Bbb C^*=\Bbb C\backslash 0$. The
generators of the Lie algebra $\sLtwo$ have the form

$$ L_{-1}=z, L_0=z\frac{\partial}{\partial z}-t\frac{\partial}{\partial t},
L_1=z(\frac{\partial}{\partial z})^2-2t\frac{\partial^2}{\partial t\partial
z}.\tag 5$$

The highest vector $v_h$ of the weight $h$ has the form $t^{-h}$. The second
realization is one in the Fock space over the Laguerre deformation of the
complex disk$^8$. The model space is the same. the generators of the Lie
algebra $\sLtwo$ have the form

$$L_{-1}=z, L_0=z\frac{\partial}{\partial z}-t\frac{\partial}{\partial t},
L_1=z(\frac{\partial}{\partial z})^2-2t\frac{\partial^2}{\partial t\partial
z}+t^2\frac{\partial}{\partial t}.\tag 6$$

The highest vector $v_h$ of the weight $h$ has the form $t^{-h}F(-h,2-2h;
tz)$, where $F(a,b;u)$ is the degenerate hypergeometric function. Other
realizations of the model should be found in the Ref.8. It should be mentined
that the realizations (5) and (6) are ones in the Fock spaces over the
manifolds, on which $\sLtwo$ acts$^8$. Therefore, the structures of
co--commutative coalgebras in the model are defined. The first structure is
rather simple:

$$\Delta(t^{\lambda}z^k)=\sum_{i+j=k}\int(t^\rho
z^i)\otimes(t^{\lambda-\rho}z^j)\,d\rho$$
but the second is interesting enough. In the realization (6) it has the same
form as the first structure in the realization (5) but now the vectors
$t^\lambda z^k$ do not generate the Verma module $V_\lambda$, so the formulas
for the comultiplication contain all Clebsch--Gordan coefficients for
$\sLtwo$.

\redefine\span{\operatorname{span}}
Now we shall be intereted in the hidden symmetries in the model. The ideology
of Res--constructions of the models was proposed by I.M.Gelfand and
A.V.Zelevinsky$^9$. Res--construction in its general form means that the model
of the representations of the Lie algebra $\frak g$ is obtained as the
restriction of an irreducible representation of an associative algebra, which
contains the universal envelopping algebra $\Cal U(\frak g)$ of the Lie
algebra $\frak g$. If one has the model of the representations of a Lie
algebra $\frak g$ and wants to obtain it by Res--construction he has to look
for the hidden symmetries in the model, which form the necessary associative
algebra together with the generators of the Lie algebra $\frak g$ $^{10}$.
Let's denote $\Cal F=\span(t^\lambda dt^\mu)$. Then the representation of the
Lie algebra $\sLtwo$ in the model of the Verma modules over it can be extended
to the action of the associative algebra $\Cal U(\sLtwo)\ltimes\Cal F$ $^8$.
The operator $T$, which represents the element $t$ of $\Cal F$ in the
realization (6) has the form

$$ T=t-\frac{\partial}{\partial z}.\tag 7$$

The operator $B_\mu$, which represents the element $(dt)^{-\mu}$ of $\Cal F$
has the form$^{11}$

$$ B_\mu=t^{-\mu}F(\mu,2+\frac{\partial}{\partial t};tz),\tag 8$$

where $F(a,b;u)$ is the normalized degenerate hypergeometric function

$$F(a,b;u)=\sum_{k\ge
0}u^k\frac{\Gamma(b-a)\Gamma(a+k)}{k!\Gamma(a)\Gamma(b+k)}.\tag 9$$

\define\ad{\operatorname{ad}}
The operator $B_{-1}$ is equal to $t^2\frac{\partial}{\partial t}+2t-t^2z$.
The operator $\nabla$ in the model of the Verma modules over the Lie algebra
$\sLtwo$ is called a connection $^8$ iff

$$ (\ad(L_n)-(n+1)(-T)^n)\nabla=0; [\nabla,f(T)]=f'(T).\tag 10$$

The operator $\nabla_h$, defined as
$\nabla_h(f(T)L^k_{-1}v_h)=f'(T)L^k_{-1}v_h$, is a connection. As it was shown
in Ref.8 an arbitrary connection in the model of the Verma modules over the
Lie algebra $\sLtwo$ coincides with one of the connections $\nabla_h$. Let $X$
be an operator in the model of the Verma modules over the Lie algebra
$\sLtwo$, which commutes with all degrees of the operator $T$. The value of
the operator $X$ in the point $u$ of the complex plane is the operator
$X(u;\nabla_h)=:\exp((T-u)\nabla_h):X$. The set of the operators
$B_{\mu}(u;\nabla_h)$ form the vertex operator fields in the model of the
Verma modules over the Lie algebra $\sLtwo$. The explicit formulas for them
are written in Ref.12. As it was shown in Ref.8 the set of the operator fields
$z^jB_\mu(u;\nabla_h)$ is closed under the operator product expansion. The
corresponding QPFT--operator algebra is denoted by $\Vrt(\sLtwo)$ and is
called the algebra of vertex operators for the Lie algebra $\sLtwo$.

Let's consider the following linear operator in $\Vrt(\sLtwo)$:

$$M_a:z^jB_\mu(u;\nabla_h)\longmapsto z^jB_{\mu+a}(u;\nabla_h).\tag 11$$

\bf Theorem 1\rm: For all $A$ and $B$ from $\Vrt(\sLtwo)$ the following
relation holds

$$A\cdot M_a(B)=M_a(A\cdot B).\tag 12$$

Proof. Such relation should be verified on the basis of $\Vrt(\sLtwo)$.
Namely,

$$\align
&z^iB_\lambda(u;\nabla_h)M_a(z^jB_\mu(v;\nabla_g))=z^iB_\lambda(u;\nabla_h)\cdot
z^jB_{\mu+a}(v;\nabla_g)=\\
&z^i
B_\lambda(u;\nabla_h)z^jB_\mu(v;\nabla_g)B_a=M_a(z^iB_{\lambda}(u;\nabla_h)\cdot
z^jB_\mu(v;\nabla_g)).
\endalign
$$

\define\Fun{\operatorname{Fun}}
So the QPFT--operator algebra $\Vrt(\sLtwo)$ admits a representation in the
space $\Fun(\Bbb R,\Bbb C[z])$, whose elements $f(h,z)$ are the classes $$\int
f(h,z)B_\lambda(u;\nabla_h)\,dh\,/\,\text{(right multiplication on
$B_\lambda$)}.$$

As it was shown in Ref.8 the QPFT--operator algebras are just the same as the
QPFT--operator subalgebras in $\Mat_N(\Vrt(\sLtwo))$. Thus, the investigation
of the fusion in the quantum projective field theory is reduced to an
investigation of algebras of matrices with coefficients in the algebra of
vertex operators for the Lie algebra $\sLtwo$. Another approach to the fusion
in the quantum projective field theory is presented in the following
paragraph.

\ \newline

\subhead III. AN ALTERNATIVE APPROACH TO THE FUSION IN THE QUANTUM PROJECTIVE
FIELD THEORY: SHIFT ALGEBRAS\endsubhead

\ \newline

\define\aff{\operatorname{aff}}
Let's consider the space $T=\Bbb C[x,x^{-1},y]$ where the variables $x$ and
$y$ do not commute: $x^mP(y)=P(y-m)x^m$. The space $T$ maybe considered as the
algebra of weighted shift operators on $\Bbb Z$ (cf.Ref.13). One should
mention that $[x,y]=x$ so $T$ is a representation space for the Lie algebra
$\aff$ of affine vector fields on $\Bbb R$. Such representation maybe extended
to a representation of the Lie algebra $W_1$ of formal vector fields on $\Bbb
R$ by the adjoint action of the elements

$$a_{-1}=x, a_0=y, a_1=x^{-1}y(y-1),\ldots a_n=x^{-n}y\ldots(y-n),\ldots\tag
13$$

It should be marked that $T=\Bbb C[x,x^{-1}]\otimes\span(a_0,a_1,a_2,\ldots
a_n,\ldots)$ and

$$a_ia_j=xa_{i+j+1}+(i+1)a_{i+j}.\tag 14$$

The elements $a_{-1},a_0,a_1$ generate the Lie algebra $\sLtwo$. Let's
consider the authomorphisms $H_h$ of $T$

$$H_h(a)=x^{-h}ax^h,\quad h\in\Bbb R,\tag 15$$
which are internal for $z\in\Bbb Z$. Let also denote

$$a_n(h)=H_h(a_n).\tag 16$$

Then

$$\align a_{-1}(h)=x, &a_0(h)=y+h, a_1(h)=x^{-1}(y+h)(y+h-1),\ldots\tag 17\\
&a_n(h)=x^{-n}(y+h)\ldots(y-n+h),\ldots\endalign$$
so that $T=\Bbb C[x,x^{-1}]\otimes(a_0(h),a_1(h),a_2(h),\ldots)$ and

$$a_i(h)a_j(h)=xa_{i+j+1}(h)+(i+1)a_{i+j}(h).\tag 18$$

The elements $a_i(h)$ ($i=-1,0,1$) generate the Lie algebra $\sLtwo$. One
should mention that

$$H_hH_{h'}=H_{h+h'}\tag 19$$
so that $H_h=\exp(hd)$. Let's consider the operator $\xi:T\mapsto T$ such that

$$\xi:a\mapsto x\frac{\partial}{\partial x}a.\tag 20$$

Then $\xi$ is a derivation of $T$. Moreover, $\xi$ commutes with all
$H_h$ and, therefore, with $d$. Let's extend the algebra $T$ by the variable
$d$ to the algebra $T'$ such that $[x,y]=x$, $[x,d]=0$, $[d,y]=1$; $\xi$ acts
in $T'$ by derivations. Let's also consider the space $\tilde T'$ in which we
admit formal series by variables $x,x^{-1},y,d$.

\bf Definition 4\rm: An element $v_\mu\in\tilde T'$ is called basic of the
weight $\mu$ iff

$$(\ad(a_n)-(-1)^n(\xi+n\mu)v_\mu=0.\tag 21$$

The elements $x^nv_\mu$, $n\ge 0$ generate the family of a basic element
$v_\mu$.

We shall denote the subspace of $\tilde T'$ of the elements from $\Bbb
C[[x,x^{-1},y]]H_h$ by $T'_h$.

\bf Theorem 2\rm: There exists the unique up to a multiple basic element of
the weight $\mu$ in the space $\tilde T'_h$, which will be denoted by
$v_{\mu,h}$. The basic element $v_{\mu,h}$ has the form

$$ v_{\mu,h}=s^+(x,y;\mu,h)H_h=H_hs^-(x,y;\mu,h),\tag 22$$

($x$ stands on the left side from $y$)

$$\align
s^+(x,n;\mu,h)&=s^+_n(x;\mu,h)\tag 23A\\
s^-(x,n;\mu,h)&=s^-_n(x;\mu,h)\tag 23B
\endalign
$$

$$\align
s^+_{n+1}(x)&=(s^+_n(x))'+(1-a^+/x)s^+_n(x)\tag 24A\\
s^-_{n+1}(x)&=(s^-_n(x))'+(1-a^-/x)s^-_n(x)\tag 24B
\endalign
$$

$$ a^+=a^-=\mu-h$$

$$\align
&(s^+_n)'''+A^+_n(x)(s^+_n)''+B^+_n(x)(s^+_n)'+C^+_n(x)s^+_n=0\tag 25A\\
&(s^-_n)'''+A^-_n(x)(s^-_n)''+B^-_n(x)(s^-_n)'+C^-_n(x)s^-_n=0\tag 25B
\endalign
$$

$$ s^+_n(x)\sim_{x\to 0}x^{-n}, s^-_n(x)\sim_{x\to 0}x^{-n}$$

$$\align
A^+_n(x)&=2+(2+b^+_n-a^+)/x\\
B^+_n(x)&=1+(2+b^+_n-a^++c^+)/x+(d^+_n-a^+b^+_n)/x^2\tag 26\\
C^+_n(x)&=(c^+-e^+_n)/x+(a^++d^+_n-a^+c^+)/x^2+a^+(b^+_n-d^+_n)/x^3
\endalign
$$

$$\align
A^-_n(x)&=2+(2+b^-_n-a^-)/x\\
B^-_n(x)&=1+(2+b^-_n-a^-+c^-)/x+(d^-_n-a^-b^-_n)/x^2\tag 27\\
C^-_n(x)&=(c^--e^-_n)/x+(a^-+d^-_n-a^-c^-)/x^2+a^-(b^-_n-d^-_n)/x^3
\endalign
$$

$$\align
&b^+_n=2n, c^+=\mu+h, d^+_n=n(n+1), e^+_n=n(n+1)-2h\tag 28\\
&b^-_n=2(n+h), c^-=\mu+h, d^-_n=(n+2h)(n+1), e^-_n=n(n+1)
\endalign
$$

Proof. The formulas for $v_{\mu,h}$ are obtained by the direct but tedious
computations, so they should be omitted.

The operator $\xi$ generates the 1--parametric family $\alpha_u$ ($u\in\Bbb
C^*$) of automorphisms of $\tilde T'$. It should be mentioned that the product
of two basic elements $v_1$ and $v_2$ is ill--defined, because they are the
formal series. Nevertheless, one may regularize their product by the
automorphism $\alpha_u$. The product $$(\alpha_uv_1)v_2=\sum_{w\text{---
basic}} u^{k(w)}P_w(x)w$$ is correctly defined Laurent series. Such OPE will
be called the regularization of product $v_1v_2$ (RP).

\bf Definition 5\rm: A set of families of basic elements in $\Mat_N(\Bbb
C)\otimes\tilde T'$, which is closed under the RP for all $u$, is called a
shift algebra.

\bf Theorem 3\rm: There exists a strict functor from the category of the
QPFT--operator algebras to the category of shift algebras and vice versa.

Proof. One should compare formulas (22--28) with ones for the vertex operator
fields in the model of the Verma modules over the Lie algebra $\sLtwo$ $^{12}$
and then use the results of the second paragraph.

It will be very interesting to give an interpretation of shift algberas in
terms of geometric quantization in a way analogous to Ref.14.

\bf Remark\rm: If $v_{\mu,h}$ is the basic of the weight $\mu$, then the
operator field $u^a\alpha_u(v_{\mu,h})$, where $a=a^-=a^+$, is
$\sLtwo$--primary.

\bf Remark\rm: (on the hidden additive weight of the $\sLtwo$--primary
fields).
{}From the fact that $\tilde T'_h\tilde T'_g=\tilde T'_{h+g}$ the additivity of
the hidden quantum number $h$ of the $\sLtwo$--basics $v_{\mu,h}$ follows.

\ \newline

\subhead IV. PROJECTIVE \"W--ALGEBRAS
\endsubhead

\ \newline

One of the most intriguing object of the quantum conformal field theory is
W--algebra $^{15}$. In this paragraph we inwestigate objects, which are
similar to W--algebras in several properties but differ in others (cf.Ref.16)
so they will be called projective \"W--algebras ("double \"u--algebras").

\bf Definition 6\rm: Let $H$ be an arbitrary QPFT--operator algebra. The
associative algebra generated in the space $H$ by operators $a_{t,a}$

\define\W{\text{\"W}}
$$\int a_{t,a}u^t\,dt=V_a(u),\tag 29$$
where $\{V_a(u)\}$ is the set of all primary fields in the QPFT--operator
algebra $H$, is called the associated projective \"W--algebra and is denoted
by $\W(H)$.

As follows from the results of Ref.8 each projective \"W--algebra is a
subalgebra in $\Mat_N(\W_{\proj})$, where $\W_{\proj}=\W(\Vrt(\sLtwo))$ is the
universal projective \"W--algebra (cf.Ref.17).

\bf Theorem 4\rm: The universal projective \"W--algebra $\W_{\proj}$ is the
algebra of the linear operators in the model of the Verma modules over the Lie
algebra $\sLtwo$. Its generators $a_{t,\mu,h}$ have the form

$$\align
a_{t,\mu,h}&=\bigoplus_k a^k_{t,\mu,h},\qquad
a^k_{t,\mu,h}:V_{h-t-\mu+k}\longmapsto V_h\tag 30\\
a^k_{t,\mu,h}&=z^kF_k(z\frac{\partial}{\partial z};\mu,h-t-\mu+k,h),\qquad k\ge
0\tag 31\\
a^k_{t,\mu,h}&=(\frac{\partial}{\partial z})^kF_k(z\frac{\partial}{\partial
z};\mu,h-t-\mu+k,h),\qquad k\le 0,\endalign$$

where the functions $F_k(x;\mu,g,h)=\sum_j F_{kj}(\mu,g,h)\delta(x-j)$ obey
the difference hypergeometric equations

$$P_{\pm}(x)\Delta^2F+Q_{\pm}(x)\Delta F+R_{\pm}F=0\tag 32$$
the signs correspond to the positive or negative $k$,

$$\align
P_+(x)&=(x+k+2)(x+k+2h+1)\tag 33\\
P_-(x)&=(x+2)(x+2h+k+1)\\
Q_+(x)&=2(k+1+h-g)x+(k+2)(k+1+2h)-2g+\\&(k+h-g+\mu)(k-\mu+h-g+1)\\
Q_-(x)&=2(1+h-g)x+2hk+2(1+2h-g)+\\&(k+h-g+\mu)(k-\mu+h-g+1)\\
R_+&=(k+h-g+\mu)(k+1-\mu+h-g)\\
R_-&=(k+h-g+\mu)(k+1-\mu+h-g)\endalign
$$
the functions $F_k$ are connected by the following equations

$$\align
F_k(x)&=(k-\mu+h-g)^{-1}\Delta F_{k-1}(x),\quad k\ge 0\tag 34\\
F_k(x)&=(k-\mu+h-g)^{-1}((x+1)F_{k-1}(x+1)-(x+k)F_{k-1}(x)),\quad k\le 0.
\endalign
$$

Proof. The theorem follows from the explicit formulas for the vertex operator
fields in the model of the Verma modules over $\sLtwo$ $^{12}$.

\ \newline

\subhead V. BRAIDING IN THE QUANTUM PROJECTIVE FIELD THEORY: PROJECTIVE
ZAMOLODCHIKOV ALGEBRAS
\endsubhead

\ \newline

The most interesting fact is that the braiding in the quantum projective field
theory is naturally described in the same terms as fusion, i.e. in terms of
vertex operator fields in contrast to the quantum conformal field theory,
where there are no any universal method for a description of the braiding.

As it was shown in Ref.8 the Sklyanin--Faddeev--Zamolodchikov
relations$^{2,18}$ for the vertex operator fields $B_\lambda(u;\nabla_h)$
hold:

$$B_{\lambda_1}(u;\nabla_{h_1})B_{\lambda_2}(v;\nabla_{h_2})
=\sum(R_{\proj}(u-v)^{\lambda_3\lambda_4 h_3h_4}_{\lambda_1\lambda_2 h_1h_2}
B_{\lambda_3}(v;\nabla_{h_3})B_{\lambda_4}(u;\nabla_{h_4}).\tag 35$$

The R--matrix $R_{\proj}(u)$ obeys the quantum Yang--Baxter equation
$^{2,19}$:

$$R_{\proj}(v-u)^{\delta\varepsilon}_{\alpha\beta}R_{\proj}(v)^{\zeta\eta}_{\varepsilon\gamma}
R_{\proj}(u)^{\theta\varkappa}_{\delta\zeta}=R_{\proj}(u)^{\delta\varepsilon}_{\beta\gamma}
R_{\proj}(v)^{\theta\zeta}_{\alpha\delta}R_{\proj}(v-u)^{\varkappa\eta}_{\zeta\varepsilon}\tag
36,$$

where
$\alpha,\beta,\gamma,\delta,\varepsilon,\zeta,\eta,\theta,\varkappa\in\Bbb
R^2$.

\bf Remark\rm:

$$(R_{\proj}(u))^{\lambda_3\lambda_4 h_3h_4}_{\lambda_1\lambda_2 h_1h_2}\ne 0
\text{ only if } h_1=h_3.\tag 37$$

\bf Theorem 5\rm:

$$ (R_{\proj}(u))^{\lambda_3\lambda_4+\mu h_3h_4}_{\lambda_1\lambda_2+\mu
h_1h_2}=(R_{\proj}(u))^{\lambda_3\lambda_4 h_3h_4}_{\lambda_1\lambda_2
h_1h_2}.\tag 38$$

Proof of the theorem is analogous to one of the theorem 1.

\bf Definition 7\rm:$^{18,19}$ Zamolodchikov algebra, correspondinig to the
R--matrix $R(u)^{ij}_{kl}$, obeying the quantum Yang--Baxter equation, is the
set of formal generators $A_i(u)$ such that

$$A_i(u)A_j(v)=R(u-v)^{kl}_{ij}A_k(v)A_l(u).\tag 39$$

Two Zamolodchikov algebras $\Cal A=\{A_i(u)\}$ and $\Cal A'=\{A'_i(u)\}$ will
be called equivalent iff the corresponding R--matrices are equivalent, i.e.

$$R'(u)^{kl}_{ij}=R(u)^{mn}_{pr}C^k_mC^l_nC^p_iC^q_j$$
for some invertible matrix $C^a_b$.

By the representation of the Zamolodchikov algebra we mean the representation
of formal generators $A_i(u)$ by the analytic operator fields so that the
product $A_i(u)A_j(v)$ exists if $|u|\gg|v|$ and maybe analytically extended
to $\Bbb C^2$. After the analytic continuation the relation (39) should hold.

\bf Theorem 6\rm: The set of the primary vertex operator fields in the
QPFT--operator algebra $\Vrt(\sLtwo))$ is the Zamolodchikov algebra.

\bf Remark\rm: It is an interesting problem to formulate the Cherednik--type
representation $^{20}$ of the R--matrix $R_{\proj}(u)$ and the Zamolodchikov
algebra of the theorem 6.

\bf Definition 8\rm: The Zamolodchikov algebra $\Cal A=\{A_i(u)\}$ is called
the projective Zamolodchikov algebra iff there exists the linear space $H$,
which can be expanded in the direct sum or the direct integral of the Verma
modules over the Lie algebra $\sLtwo$, and the representation of $\Cal A$ in
$H$ by the primary operator fields:

$$[L_k,A_i(u)]=(-u)^k(u\frac{\partial}{\partial u}+(k+1)s_i)A_i(u).$$

It is an interesting problem to describe the class of projective Zamolodchikov
algebras in the internal terms.

\bf Hypothesis\rm: Each Zamolodchikov algebra maybe imbed into a projective
Zamo\-lodchikov algebra.

By the representation of the projective Zamolodchikov algebra we mean a
representation of its generators by the primary fields. One of the
characteristics of a representation of the projective Zamolodchikov algebra is
its weight $(s_1,\ldots s_n)$, where $s_i$ are the weights of the primary
operator fields, corresponding to the generators $A_i(u)$.

\bf Lemma\rm: Let $H$ be a linear space, which can be expanded into a direct
sum or a direct integral of the Verma modules over the Lie algebra $\sLtwo$.
$\Cal V=\{V_i(u)\}$ is the set of primary operator fields in $H$. The
QFT--operator algebra generated by $V_i(u)$ is a QPFT--operator algebra.

Proof. One should mention that the space $\bigoplus <z^jV_a(u)>$, where
$\{V_A(u)\}$ is the set of all primary fields in $H$, is closed onder the OPE
and so it is a QPFT--operator algebra. Also if the OPE of two primary fields
contains a secondary of the third one then it contains this primary
field$^{21}$. So the QFT--operator algebra generated by $\{V_i(u)\}$ is a
QPFT--operator subalgebra in $\bigoplus <z^jV_a(u):V_a(u)\text{ --- all
primary operator fields}>$.

\bf Theorem 7\rm: Let $\Cal A$ be a projective Zamolodchikov algebra
represented in $H$. The QFT--operator algebra generated by $A_i(u)$ is a
QPFT--operator algebra.

Proof. The statement of the theorem is a sequence of the previous lemma.

\define\QPFT{\operatorname{QPFT}}
\bf Definition 9\rm: Let $\Cal A$ be a projective Zamolodchikov algebra
represented in $H$. The QPFT--operator algebra constructed in the previous
theorem will be called the envelopping QPFT--operator algebra of $(\Cal A,H)$
and will be denoted by $\QPFT(\Cal A,H)$.

It is an interesting question whether exists the universal envelopping
QPFT--operator algebra of a projective Zamolodchikov algebra $\Cal A$.

Different representations of the fixed projective Zamolodchikov algebra may
have the same envelopping QPFT--operator algebras, therefore, a natural
problem of the representation theory of projective Zamolodchikov algebras is
to describe the possible envelopping QPFT--operator algebras.

\bf Definition 10\rm: A QPFT--operator algebra is called braiding admissible
QPFT--operator algebra (b.a.QPFT--operator algebra) iff there exists a basis
of primary operator fields in it, which forms a projective Zamolodchikov
algebra.

It is an interesting question to find the conditions, which extract the
b.a.QPFT--operator algebras from all QPFT--operator algebras.

One has a functor from the category of representations of the projective
Zamolodchikov algebras to the category of b.a.QPFT--operator algebras. What
are the projective Zamolodchikov algebras, which have a common
b.a.QPFT--operator algebra?

Now we shall describe an important object connected with representations of
projective Zamolodchikov algebras. It is analogous to spherical functions in
the group representation theory$^{22}$ and so will be called a spherical
correlation function.

\bf Definition 11\rm: Let $\Cal A$ be a projective Zamolodchikov algebra
represented in the linear space $H=\oplus V_a$, where $V_a$ is a Verma module
over $\sLtwo$ with a highest vector $v_a$ and a highest weight $h_a$. The
spherical correlation functions of generators $A_{i_1}(u_1),\ldots
A_{i_n}(u_n)$ are

$$<A_{i_1}(u_1)\ldots A_{i_n}(u_n)>_{ab}=<v_a|A_{i_1}(u_1)\ldots
A_{i_n}(u_n)|v_b>.\tag 40$$

\bf Theorem 8\rm: A representation of a projective Zamolodchikov algebra can
be constructed from the spherical correlation functions

$$<A_i(u)>_{ab}.\tag 41$$

Proof. It should be mentioned that a representation of a projective
Zamolodchikov algebra can be characterized by its weight $(s_1,\ldots s_n)$
and the set of the constants

$$c_{k,ab}=\Proj_{V_b}(\left.(A_k(u)\right|_{V_a})/V_{s_k}(u;h_a,h_b)$$
where $V_s(u;h,g)$ is the standard primary operator field of the weight $s$
acting from the Verma module $V_h$ to the Verma module $V_g$ $^{8,12}$. Let us
mark that

$$<v_a|V_{s_k}(u;h_a,h_b)|v_b>=u^{-s_k+(h_b-h_a)}.$$
So one can reconstruct the weight $s_k$ and the constant $c_{k,ab}$ from the
correlation function $<A_k(u)>_{ab}$.

Let's consider the matrix generators $E_{ab}$. Using them one can collect
spherical correlation functions in the following expressions

$$<A_{i_1}(u_1)\ldots A_{i_m}(u_m)>:=\sum_{a,b}<A_{i_1}(u_1)\ldots
A_{i_m}(u_m)>_{ab} E_{ab}.$$

\define\Tr{\operatorname{Tr}}
\bf Definition 12\rm: Let $\Cal A$ be a projective Zamolodchikov algebra
represented in $H$ by primary operator fields $A_i(u)$. The correlation
character of generators\linebreak $A_{i_1}(u_1),\ldots A_{i_m}(u_m)$ is

$$ \Tr<A_{i_1}(u_1)\ldots A_{i_m}(u_m)>.\tag 43$$

It should be mentioned that as a rule the trace does not exist in the na\"\i
ve sense so one should consider it either as a trace in a suitable von Neumann
algebra or as a generalized function (using some smoothing by $u_k$), it is
possible also to regularize the trace in some way, f.e. by an insertion of the
multiple $q^{L_0}$.

It is an interesting question whether a representation of the projective
Zamolodchikov algebra can be reconstructed from the correlation character.

\ \newline

\subhead VI. AN ALTERNATIVE APPROACH TO THE BRAIDING IN THE QUANTUM PROJECTIVE
FIELD THEORY: INFINITE DIMEN-\linebreak SIONAL PARAMETRIC R--MATRIX
$R^*_{\proj}(u)_{h,h'}$\endsubhead

\ \newline

Let's denote $\alpha_u(v_{\lambda,h})$ by $v_{\lambda,h}(u)$.

\bf Theorem 9\rm: For all $h$ and $h'$ the forllowing relation holds
(cf.Ref.6,23):

$$v_{\lambda,h}(u)v_{\lambda',h'}(v)=
{(R^*_{\proj}(u/v))_{h,h';}}^{\mu\mu'}_{\lambda\lambda'}
v_{\mu,h}(v)v_{\mu',h'}(u),\tag 44$$

the matrix $R^*_{\proj}(u)_{h,h'}$ obeys the following Yang--Baxter--type
equation

$$\align
{R^*_{\proj}(u/v)_{h,g;}}^{\delta\varepsilon}_{\alpha\beta}
&{R^*_{\proj}(u/w)_{g,f;}}^{\zeta\eta}_{\varepsilon\gamma}
{R^*_{\proj}(v/w)_{h,g;}}^{\theta\varkappa}_{\delta\zeta}=\\
&{R^*_{\proj}(v/w)_{g,f;}}^{\delta\varepsilon}_{\beta\gamma}
{R^*_{\proj}(u/w)_{h,g;}}^{\theta\zeta}_{\alpha\delta}
{R^*_{\proj}(u/v)_{h,f;}}^{\varkappa\eta}_{\zeta\varepsilon}.
\tag 45\endalign$$

Proof. the theorem follows from the fact that for all $h,h',u,v$ the operators
$v_{\lambda,h}(v)v_{\lambda',h'}(u)$ form a basis in the space $\tilde
T'_{h+h'}$. The dependence of the R--matrix on the quotient $u/v$ follows from
properties of the transformations of fields under the action of $\xi$.

\ \newline

\subhead VII. PROJECTIVE G--HYPERMULTIPLETS\endsubhead

\ \newline

\bf Definition 13\rm: Let $G$ be a group. A QPFT--operator algebra will be
called a projective $G$--hypermultiplet iff $G$ acts in the space $H$ of the
algebra by automorphisms and $G$ commutes with $\sLtwo$.

It can be easily shown that the spaces of the highest vectors of a fixed
highest weight in a projective $G$--hypermultiplet form representations of
$G$. Such representations will be called multiplets of projective
$G$--hypermultiplet.

Let's construct several examples of projective hypermultiplets.

\bf Definition 14\rm: Let $\frak g$ be a Lie algebra, $G$ --- a group of its
automorphisms. Let $t$ be a representation of $\frak h$, $g$ --- an element of
$G$. We'll denote by $t^g$ the representation of $\frak h$, which is defined
as follows

$$A\longmapsto t(g(A)).$$
A representation $t$ of $\frak h$ will be called $G$--stable iff $t^g$ are
equivalent to each other for all $g$.

\define\PSltwo{\operatorname{\widetilde {PSl}}(2,\Bbb R)}
\bf Lemma\rm: Let $\frak h$ be a Lie algebra with $\PSltwo$ as an automorphism
group. Then the space of an arbitrary $\PSltwo$--stable representation $t$ of
$\frak h$ admits a structure of $\PSltwo$--representation $T$ so that
$t(g(A))=T(g)t(A)T(g^{-1})$.

\bf Definition 15\rm: (Cf.Ref.24) Let $\frak h$ be an arbitrary Lie algebra
with basis $X^m$ and commutation relations

$$[X^m,X^n]=c^{mn}_pX^p\tag 46$$
then the truncated current algebra $^t\hat{\frak h}$ over $\frak h$ is the Lie
algebra with basis $T^m_r$, $r\ge 0$, and the commutation relations

$$[T^m_r,T^n_s]=c^{mn}_lT^l_{r+s}.\tag 47$$

One should mention that an action of $\sLtwo+\frak h^{\Bbb C}$ in $^t\hat{\frak
h}^{\Bbb C}$ defined as

$$ L_k(T^m_r)=rT^m_{r+k},\quad X^m(T^n_r)=c^{mn}_lT^l_r\tag 48$$
maybe extended to a representation of $\PSltwo\times H$ in $^t\hat{\frak
h}^{\Bbb C}$ by automorphisms.

\bf Theorem 10\rm: Let $t$ be an arbitrary $\PSltwo\times H$--stable
representation of $^t\hat{\frak h}^{\Bbb C}$ and suppose that the
corresponding representation $T$ of $\sLtwo$ belongs to the category $\Cal O$
$^{25}$. Then the operator fields (currents)

$$ T^m(u)=\sum_{r\ge 0} T^m_ru^{-1+r}+\text{ regular terms }\tag 49$$
generate a projective $H$--hypermultiplet.

Proof. The theorem folows from the lemma before the theorem 7 and the fact
that an induced action of $H$ on the QPFT--operator algebra, generated by the
currents $T^m(u)$, preserves the structure of the QFT--operator algebra and
$\sLtwo$--action.

Let' s now consider another example of projective $G$--hypermultiplet.

\define\Aut{\operatorname{Aut}}
\define\Ind{\operatorname{Ind}}
\bf Theorem 11\rm: Let $\Cal A$ be a projective Zamolodchikov algebra with a
group $G$ of its automorphisms and $H$ --- an arbitrary space, where $\Cal A$
is represented. Let $K$ be a subgroup of $G$ such that
$K\subseteq\Aut(\QPFT(\Cal A,H))$. Then the QPFT--operator algebra $\QPFT(\Cal
A,\hat H)$, where $\hat H=\Ind^G_K(H)$, is a b.a.projective
$G$--hypermultiplet.

\ \newline

\subhead VIII. CONCLUSIONS\endsubhead

\ \newline

Thus, various structures of the quantum projective field theory related to
fusion and braiding were explored. projective Zamolodchikov algebras, their
representations, spherical correlation functions and correlation characters,
envelopping QPFT--operator algebras, projective \"W--algebras, shift algebras,
infinite dimensional R--matrices, projective $G$--hypermultiplets are among
them. various approaches to fusiona and braiding are presented.

It is proved (in the formalism of shift algebras) that $\sLtwo$--primary
fields are characterized by their projective weights and a hidden
quantum number discovered in the paper --- the hidden additive weight.

\ \newline

\subhead IX. ACKNOWLEDGEMENTS\endsubhead

\ \newline

The author thanks A.Yu.Morozov, A.A.Rosly, A.S.Losev,
M.A.Olshanetsky,\linebreak A.Marshakov, A.Gerasimov and other participants of
the Seminar on Quantum field Theory in ITEP (Moscow) for an attention and
remarks.

\Refs
\roster
\item"$^1$" E.Witten, "Physics and geometry", Rep. 100th Anniversary A.M.S.,
1988; L.D.Faddeev, Preprint SPhT 82/76, CEN Saclay, 1982; A.A.Belavin,
A.M.Polyakov, and A.B.Zamolod\-chikov, Nucl. Phys. B241, 333 (1984);
A.B.Zamolod\-chikov, and Al.B.Zamolod\-chikov, Preprints ITEP 112-89, 173-89,
31-90.
\item"$^2$" E.K.Sklyanin, and L.D.Faddeev, Dokl. Akad. Nauk SSSR 243(6) 1430
(1978).
\item"$^3$" N.Yu.Reshetikhin, LOMI Preprint E-4-87, 1988; N.Yu.Reshetikhin,
L.A.Takhtajan, and L.D.Faddeev, Algebra Anal. 1(1), 178 (1989);
L.C.Biedenharn, "An overview of quantum groups", Proc. 18th Intern. Colloq.
Group Methods, Moscow, 1990; N.Yu.Reshetikhin, and M.A.Semenov-Tian-Shansky,
Lett. Math. Phys. 19, 133 (1990).
\item"$^4$" V.G.Drinfeld, "Quantum groups", Proc. Intern. Congr. Math.,
Berkeley, 1986; Yu.I.Manin, Preprint CRM--1561, Montreal, 1988.
\item"$^5$" G.Segal, Preprint MPI/87-85, 1988; E.Witten, Commun. Math. Phys.
113, 529 (1988); M.Kontsevich, unpublished notes, 1988; L.Alvarez-Gaume,
C.Gomez, G.Moore, and\linebreak C.Vafa, Nucl. Phys. B303, 455 (1988).
\item"$^6$" G.Moore, and N.Seiberg, Commun. Math. Phys. 123, 177 (1989).
\item"$^7$" A.Z.Patashinskii, and V.L.Pokrovskii, "Fluctuation theory of phase
transitions", Pergamon Press, 1978; R.E.Borcherds, Proc. Nat'l Acad. Sci. USA
83, 3068 (1986); I.Frenkel, J.Lepowsky, and A.Meurman, "Vertex operator
algebras and the Monster", New York, 1988; P.Goddard, "Meromorphic conformal
field theory" in "Infinite dimensional Lie algebras and groups", World
Scientific, 1989.
\item"$^8$" D.V.Juriev, Algebra Anal. 3(3), 197 (1991), Uspekhi Matem. Nauk
46(4), 115 (1991); S.A.Bychkov, and D.V.Juriev, Teor. Matem. Fiz. 97(3)
(1993).
\item"$^9$" I.M.Gelfand, and A.V.Zelevinsky, Funkt. anal. i ego prilozh.
18(3), 14 (1984).
\item"$^{10}$" D.E.Flath, and L.C.Biedenharn, Canad. Math. J. 37(4), 710
(1985).
\item"$^{11}$" D.Juriev, Lett. Math. Phys. 21, 133 (1991).
\item"$^{12}$" D.Juriev, Lett. Math. Phys. 22, 141 (1991).
\item"$^{13}$" Yu.D.Latushkin, and A.M.Stepin, Uspekhi Matem. Nauk 46(2), 85
(1991).
\item"$^{14}$" A.Yu.Alekseev, and S.L.Shatashvili, Commun. Math. Phys. 128,
197 (1990).
\item"$^{15}$" F.A.Bais, P.Bouwknegt, M.Surridge, and K.Schoutens, Nucl. Phys.
B304,
348 (1988); I.Bakas, Commun. Math. Phys. 123, 627 (1989); O.D.Ovsienko, and
V.Yu.Ovsienko, Adv. Soviet Math. 2, 221 (1991).
\item"$^{16}$" A.Bilal, Preprint PUPT--1434;
hep-th/9312108; Preprint PUPT--1446; hep-th/9401167.
\item"$^{17}$" A.Yu.Morozov, Preprint ITEP 89/148, 1989; A.O.Radul, Funkt.
anal. i ego prilozh. 25(1), 33 (1991).
\item"$^{18}$" E.K.Sklyanin, L.A.Takhtajan, and L.D.Faddeev, Teor. Matem. Fiz.
40, 194 (1979);\linebreak  A.B.Zamolodchikov, Pis'ma ZhETP 25(10), 13 (1977).
\item"$^{19}$" L.A.Takhtajan, and L.D.Faddeev, Uspekhi Matem. Nauk 34(5), 13
(1979).
\item"$^{20}$" I.V.Cherednik, Dokl. Akad. Nauk SSSR 249(5), 1095 (1979).
\item"$^{21}$" A.A.Belavin, A.M.Polyakov, and A.B.Zamolodchikov, Nucl. Phys.
B241, 333 (1984).
\item"$^{22}$" D.P.Zhelobenko, and A.I.Schtern, "Representations of Lie
groups", Moscow, Nauka, 1983.
\item"$^{23}$" G.Moore, and N.Yu.Reshetikhin, Preprint IASSNS--HEP--89/18,
1989.
\item"$^{24}$" V.G.Kac, "Infinite dimensional Lie algebras", Cambridge, 1985.
\item"$^{25}$" I.N.Bernstein, I.M.Gelfand, and S.I.Gelfand, Funkt. anal. i ego
prilozh. 5(1), 1 (1971).
\endroster
\endRefs
\enddocument